\documentclass[aps,prl,twocolumn,groupedaddress]{revtex4-1}
\usepackage{graphicx}
\usepackage{breqn}

\begin{document}

\title{Time- and Site- Resolved Dynamics in a Topological Circuit}
\author{Jia Ningyuan}
\author{Clai Owens}
\author{Ariel Sommer}
\author{David Schuster}
\author{Jonathan Simon}
\affiliation{Physics Department and James Franck Institute at the University of Chicago}
\date{\today}

\begin{abstract}
From studies of exotic quantum many-body phenomena~\cite{RevModPhys.71.S298} to applications in spintronics and quantum information processing~\cite{PhysRevLett.104.040502,RevModPhys.80.1083}, topological materials are poised to revolutionize the condensed matter frontier and the landscape of modern materials science. Accordingly, there is a broad effort to realize topologically non-trivial electronic and photonic materials for fundamental science as well as practical applications. In this work, we demonstrate the first simultaneous site- and time- resolved measurements of a time reversal invariant topological band-structure, which we realize in a radio frequency (RF) photonic circuit. We control band-structure topology via local permutation of a traveling wave capacitor-inductor network, increasing robustness by going beyond the tight-binding limit. We observe a gapped density of states consistent with a modified Hofstadter spectrum~\cite{Hofstadter1976,PhysRevLett.80.3232} at a flux per plaquette of $\phi=\pi/2$. In-situ probes of the band-gaps reveal spatially-localized bulk-states and de-localized edge-states. Time-resolved measurements reveal dynamical separation of localized edge-excitations into spin-polarized currents. The RF circuit paradigm is naturally compatible with non-local coupling schemes, allowing us to implement a M\"{o}bius strip topology inaccessible in conventional systems. This room-temperature experiment illuminates the origins of topology in band-structure, and when combined with circuit quantum electrodynamics (QED) techniques~\cite{Wallraff2004},  provides a direct path to topologically-ordered quantum matter.
\end{abstract}

\maketitle

Global topological features arise in a variety of contexts from knotted vortices in classical fluids~\cite{Kleckner2013} and superfluids~\cite{Ikegami2013} to anyons~\cite{PhysRevLett.49.957} in quantum wires~\cite{1063-7869-44-10S-S29}, topological insulators~\cite{PhysRevLett.100.096407}, and  2DEGs~\cite{PhysRevLett.53.722}.   In a condensed matter context, such ``topologically protected'' properties include single-particle features of the band-structure and many-particle ground-state degeneracies, with the latter typically emerging from the former in conjunction with strong interactions. To explore the nature of topologically-derived material properties, it is desirable to develop materials that not only support conserved topological quantities, but which may be precisely produced, manipulated, and probed. The aim, then, is to realize material test-beds that marry favorable coherence properties, strong interactions, and topologically non-trivial single-particle dynamics.

Meta-materials, where interaction strengths and length scales can be engineered, are a promising avenue for studying topological physics. Efforts are ongoing to produce the requisite topological single-particle dynamics in ultracold atomic gases~\cite{Lin2009, Beeler2013,PhysRevLett.109.095301, PhysRevLett.109.095302,PhysRevLett.106.175301, Miyake2013, PhysRevLett.107.255301}, gyrotropic metamaterials~\cite{Wang2008,Wang2009}, and photonic systems~\cite{PhysRevA.84.043804,PhysRevLett.100.013904, Wang2008, Rechtsman2013a, Hafezi2011, Hafezi2013a, Khanikaev2013, Koch2010a, Mittal2014}.

In cold atomic gases, gauge fields are generated either through spatially dependent Raman-coupling of internal atomic states~\cite{Lin2009,PhysRevLett.106.175301}, or time- and space- periodic modulation of lattice tunneling rates~\cite{Jotzu2014a,Miyake2013,Aidelsburger2013}. In the optical domain, synthetic magnetic fields were realized via strain of a honeycomb lattice~\cite{Rechtsman2013b}. A Floquet topological insulator~\cite{Kitagawa2010,Lindner2011} was realized under a space-to-time mapping of an array of tunnel-coupled waveguides modulated along their propagation direction~\cite{Rechtsman2013a}. A photonic topological circuit was realized through an array of off-resonantly coupled silicon optical ring resonators~\cite{Hafezi2011,Hafezi2013a,Mittal2014}, similar to a proposal to couple Bragg stacks~\cite{PhysRevA.84.043804}. Time-reversal broken topological metamaterials have been realized in the microwave domain via a lattice of chiral magnetic (gyrotropic) resonators ~\cite{Wang2008,Wang2009}. Recently, topological invariants have been measured for an individual \cite{PhysRevLett.113.050402,Roushan2014} and pair of~\cite{Roushan2014} superconducting qubits, as well as a two-site RF network with periodic boundary conditions~\cite{PhysRevX.5.011012}.

In this work, we present and experimentally characterize the first RF circuit exhibiting a time-reversal symmetric topologically non-trivial band-structure. Our approach shares features with a number of the aforementioned proposals and experiments. Uniquely, the gauge field is realized through permuted couplings rather than phase shifts. Furthermore, we are able to temporally- and spatially- resolve the spin-filtered dynamics at the single-lattice-site level, and employ non-local couplings to realize a M\"{o}bius topology.

Our lattice may be viewed as a spin- dependent gauge field for RF photons in a network of capacitively-coupled inductors, where the spin state is encoded in two equivalent inductors on each lattice site. The simplicity of the approach points the way to straight-forward implementations of spin-orbit coupled quantum wires, fractional quantum hall systems, and proximity-coupled TI-superconductors, all within the circuit QED framework~\cite{Wallraff2004}.

\section{Engineering Topological Circuitry}

Materials that insulate in their bulk and conduct on their surfaces, topological insulators (TI) exhibit unique behaviors first observed in high-purity 2DEGs~\cite{Klitzing1980}. As in a conventional band insulator, a full valence band in a topological insulator leads to zero conductance in the bulk. The surface of such a system, however, possesses spin-filtered edge modes~\cite{PhysRevLett.95.226801} residing in  the energy-gap between valence- and conduction- bands.  These modes arise from a topological phase transition at the boundary of the topological insulator.   Within the TI, in the absence of magnetic disorder, $S_z$ is conserved and particles within the resulting spin-subbands acquire finite Berry phase when they circulate the Brillouin zone.  This results in a non-zero spin-Chern number ($C^{\uparrow \downarrow}$)~\cite{PhysRevLett.97.036808}. Wave-function continuity quantizes the Berry phase to multiples of $2 \pi$, precluding a smooth drop to zero across the boundary out of the TI, and leading (via the bulk-boundary correspondence~\cite{Hasan2010a}) to a set of mid-gap topologically protected edge modes.

There are a variety of ways to engineer topologically non-trivial band-structures in lattice models, which may be classified either as time-reversal-symmetry -conserving or -breaking. Among the time-reversal-breaking models, the simplest arises when a constant magnetic field is applied to a charged particle confined in a 2D periodic structure, as described by Hofstadter~\cite{Hofstadter1976}. The time-anti-symmetric Lorentz force is equivalent to an Aharanov-Bohm phase (flux) per plaquette $\phi=2 \pi M/N$ (for relatively prime integers $M,N$).  This breaks the intrinsic translational invariance of the lattice, resulting in an effective unit cell of size $N$ sites, and $N$ corresponding sub-bands. 

To realize magnetic-field-like physics in the absence of magnetic fields, or (as in the present work) for charge-neutral photons, one can introduce a pseudo-spin degree of freedom in analogy to the spin-Hall effect. Opposing spin states are made to experience opposing magnetic fields through spin-orbit coupling. Such models, which ``break'' time reversal symmetry oppositely for up and down spins, thus do not violate the symmetry at all. They produce two copies of the Hofstadter model, exhibiting opposite effective magnetic fields everywhere in space for the two spin states, without the need for an applied magnetic field. In the solid-state such models rely on either Dresselhaus or Rashba spin orbit couplings~\cite{Hasan2010a}, arising from atomic spin-orbit interactions and relativistic coupling to static electric fields respectively.

\begin{figure}[h!]
\includegraphics[width=3.4in]{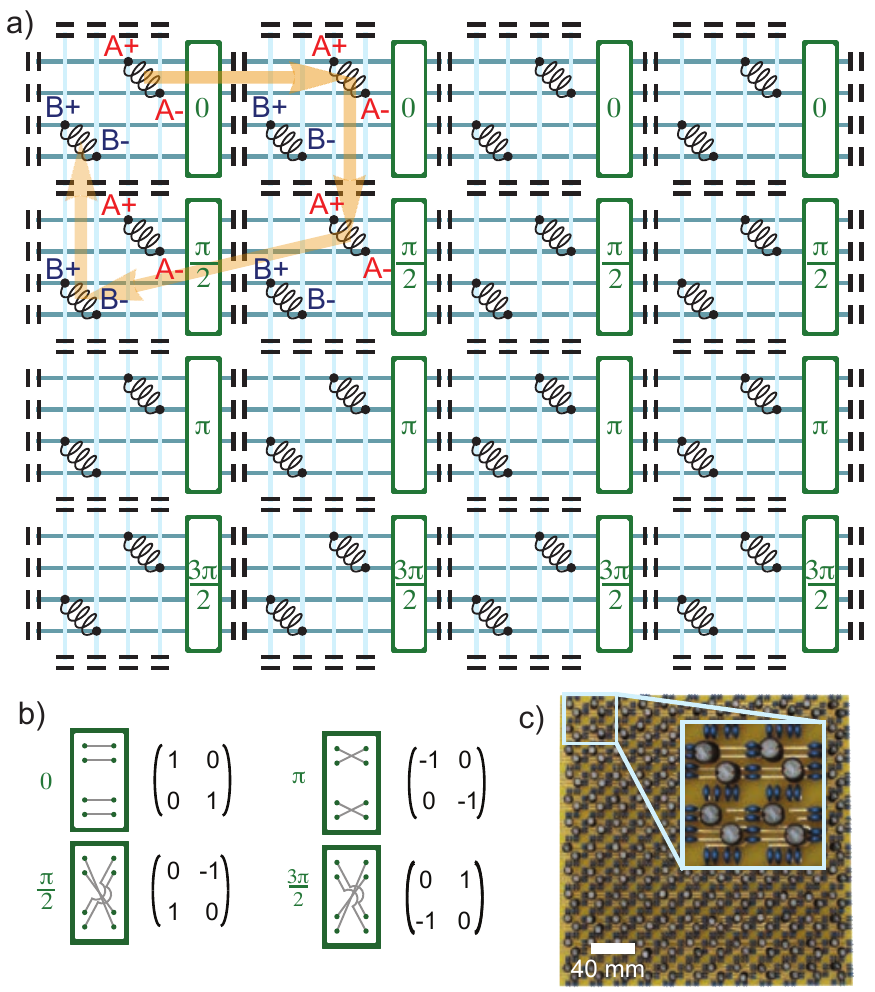}
\caption{\label{fig1}
a. Circuit topological insulator schematic. The periodic structure is formed by onsite inductors and coupling capacitors, (black) that are connected via a latticework of wires (light and dark blue lines). At each lattice site, the two inductors ``A'' and ``B'' correspond to right and left circularly polarized spins. When a photon traverses a single plaquette (indicated in orange) it accumulates a Berry phase of $\pi / 2$. The phase is induced by braiding (indicated by the green boxes and specified in b) of the capacitive couplings. b. Structure of the coupling elements between lattice sites. Each row shows one of the four rotation angles implemented by the capacitive coupling in the circuit. The rotation angle (left column) is induced by connecting inductors as shown (middle column). The corresponding rotation matrices (right column) indicate the inductors being coupled, as well as the signs of the couplings.  c. Photograph of Circuit Topological Insulator. The inductors (black cylinders) are coupled via the capacitors (blue); circuit topology is determined by the trace layout on the printed circuit board (yellow). Inset: Zoom-in on a single plaquette consisting of four adjacent lattice sites.
}
\end{figure}

In the present work we generate spin-orbit coupling through local circuit connections (Figure \ref{fig1}a,b): Two arrays of inductors provide the ``up''- and ``down''- spin RF photons, while the kinetic couplings are provided by capacitors which induce a flux per plaquette of $\phi= \pi/2 = 2 \pi /4$, eg $M=1$, $N=4$ in the corresponding Hofstadter model. As such, this flux requires a lattice with a ``magnetic'' unit cell of $N=4 \times 1$ sites in order to ensure the flux is the same on all plaquettes. This enlargement of the physical unit cell is a generic feature of meta-material implementations of magnetic fields, pointed out by Hofstadter in the context of his famous butterfly ~\cite{Hofstadter1976}- once a gauge is chosen for the Aharanov Bohm phase, it is apparent that the physical sites/couplings of the lattice must be modified (the unit cell enlarged) to affect the appropriate couplings. In our work this arises from spatially modulated permutation of capacitive couplers. In the work of Hafezi and colleagues ~\cite{Hafezi2013} this takes the form of a spatial modulation of the location of the coupling resonators.

To understand the origin of the flux in this system, we track an excitation as it traverses a single plaquette. The state of excitation on each lattice site is given by a complex vector ($V_A$,$V_B$), with $V_A$ and $V_B$ the voltages across inductors $A$ and $B$, respectively, in the frequency domain. For simplicity we define an excitation localized on $A$ and $B$ lattice sites as: $(V_A,V_B)=(1, 0)\equiv \textbf{A}$ and $(0,1) \equiv \textbf{B}$, respectively. Consider a photon which begins in the $A$ sub-lattice at site [0,0], and moves around a plaquette following the coupling topology according to Figure \ref{fig1}a,b: The capacitive couplings connect $\textbf{A} \rightarrow \textbf{A} \rightarrow -\textbf{B} \rightarrow -\textbf{B}$; the connection $\textbf{A} \rightarrow -\textbf{B}$ corresponds to a rotation of $-\pi/2$ in the $\textbf{A}$, $\textbf{B}$  space (as shown in the second row of Fig. \ref{fig1} b).

Following the same path for a $B$ excitation at site $[0,0]$ results in $A$ after a full loop. A change of basis to $\uparrow$,~$\downarrow$~$= (\textbf{A} \pm i \textbf{B})/\sqrt{2}$, reveals that after the same loop, $\uparrow$ becomes $i \uparrow$, and $\downarrow$ becomes $-i \downarrow$, which is precisely the $\pi/2$ flux per plaquette that was sought. The horizontal coupling changes from row-to-row (Figure \ref{fig1}b) with a four row period (as in the Hofstadter model with $M=1$, $N=4$), to ensure a flux of $\pi/2$ in every plaquette.

Because each spin-state propagates in a fixed direction on the edge of a topological insulator, such states simply navigate around any disorder that does not flip the spin; backscattering requires a magnetic impurity that breaks time-reversal symmetry ~\cite{PhysRevLett.95.146802}. In a photonic meta-material, impurities similarly take two forms: (1) ``non-magnetic'' disorder which is common to components (inductors and capacitors) in A- and B- sublattices, and (2) ``magnetic'' disorder which is differential for components in the sublattices. Time-reversal symmetry need not be broken to induce back-scattering in a photonic topological insulator, as photons are not protected by the Kramers degeneracy ~\cite{Hasan2010a}. Such time-symmetric ``magnetic'' disorder is possible in all meta-material topological insulators: in the silicon ring-resonator experiments ~\cite{Hafezi2013}, resonator imperfections induce backscattering, mixing right- and left-handed modes and thereby flipping spins; in the silicon waveguide Floquet topological-insulator experiments ~\cite{Rechtsman2013a}, imperfections in the waveguides scatter forward-propagating modes into their back-propagating counterparts.

Figure \ref{fig1}c shows a photograph of the actual $12 \times 12$ site topological meta-material; the black cylinders are 3.3mH wire-wound inductors, and the small blue rectangles between them are 330pF coupling capacitors. The topside of the FR4 printed circuit board (PCB) contains traces for the (spatially varying) horizontal couplings, while the bottom side contains the homogeneous vertical couplings. To characterize the RF components, the frequency response of a single LC series resonator was measured, driven by a $5 \Omega$ source to avoid shunting, indicating a $Q \approx 88$ for a $2^{\rm nd}$ order pole at 120 kHz.  The component-value disorder is found to be $< 1\%$, and lower disorder can be engineered in state-of-the-art superconducting systems~\cite{Underwood2012}, where $\approx 0.01\%$,  has been demonstrated.
The proper figure of merit for topological protection is the ratio of mean tunneling rate to the rate of loss and backscatter: the mean number of protected tunnel hops. In all current topological meta-materials, loss dominates over backscatter. We observe $\sim 20$ hops/3dB loss (see Fig. \ref{fig3}a); silicon ring-resonator implementations observe $\sim 6$ hops/3dB~\cite{Mittal2014};  Floquet waveguide implementations observe $\sim 3$ hops/3dB~\cite{Rechtsman2013a}.     

Typical (non-topological) microwave lattice experiments employ a tight binding model in which a resonator represents a site, and perturbative tunnel couplings are realized through inter-site inductors or capacitors~\cite{Underwood2012}, enhancing sensitivity to onsite disorder by the ratio of the tunnel coupling to the resonator frequency. A direct application of this approach to the topological case would employ an LC resonator at each of the A and B sub-lattice sites, with a weak inductive- or capacitive- coupling (for positive- and negative- couplings respectively) between sites. To minimize sensitivity to resonator disorder, it is advantageous to have as large a tunnel coupling as possible. Taking this idea to the extreme, we eliminate the resonators entirely by making the coupling capacitance so large that the onsite capacitor may be omitted; this corresponds to moving beyond the tight-binding limit to a scale invariant regime where the particular values L and C impact only the overall energy scale of the band-structure (other fluxes per plaquette may be accessed by including appropriately engineered A-to-A,A-to-B,B-to-A, and B-to-A couplings on every link). The sign of the tunnel coupling is then controlled by which ends of the on-site inductors are capacitively connected to one-another. We calculate numerically and show experimentally that all topological properties are preserved even in this extreme case. 

Figure \ref{fig2}b shows a numerical diagonalization (described in SI) of the circuit modes of an infinite strip with 23 lattice sites in the transverse dimension, with definite spin and longitudinal quasi-momentum $q$. The four broad bands (gray curves in Fig. \ref{fig2}b) correspond to the bulk response of the system, their breadth owing to the existence multiple transverse modes in the bulk. Spin-helicity coupled edge channels, characteristic of a topological band-structure, occupy the gaps between bulk bands. In the top-gap, $\uparrow$ excitations propagate leftward on the top edge and rightward on the bottom edge, while $\downarrow$ excitations propagate rightward on the top edge and leftward on the bottom edge. The direction of propagation may be ascertained from the slope of the energy-momentum dispersion. The locking of spin to propagation direction on each edge prevents backscattering in the absence of a spin flip disorder. As discussed in the SI, the topological character of each isolated spin-band may be formally characterized in terms of a spin-Chern number, which, for $\uparrow (\downarrow)$ bands is $+1(-1)$ for the top and bottom bands and $-2(+2)$ for the sum of the middle two bands, which touch at Dirac points and thus may not be characterized independently.

\section{Characterizing the Meta-Material}

 \begin{figure}[h!]
\includegraphics[width=3.4in]{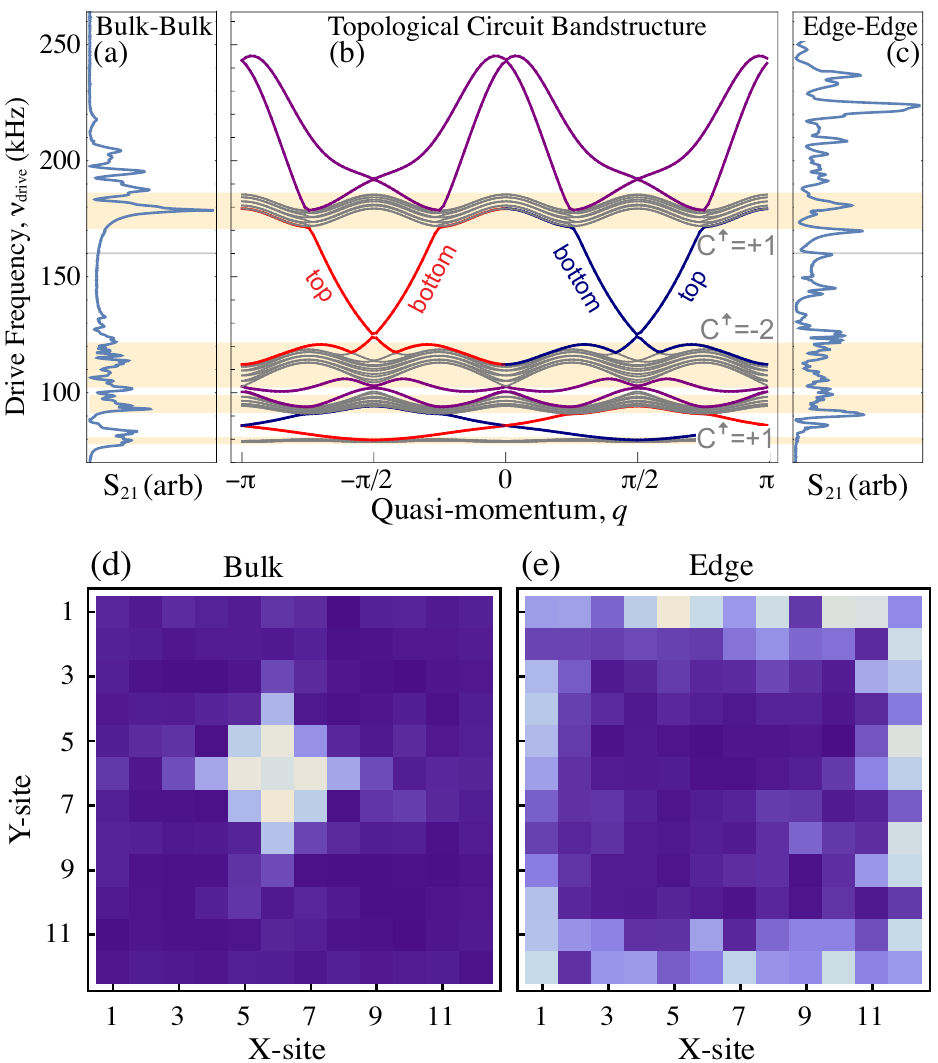}
\caption{\label{fig2}
Site-Resolved Measurement of Band-Structure and DOS. 
a. The experimentally observed coupling (linear scale) between two points in the bulk of the circuit TI (blue) with the bulk states indicated (orange bands). b. Band Structure of a Circuit TI. A strip of circuit TI with fixed boundary conditions in the transverse direction is numerically diagonalized at finite longitudinal quasi-momentum $q$, yielding four massive bulk bands (gray), and spin-orbit-locked edge states (red=$\uparrow$, blue=$\downarrow$) that reside in the bulk gap. Labels of ``top'' and ``bottom'' denote the boundary that each edge mode propagates along (with opposing group velocities $d\omega/dk$). The spin-Chern numbers ($C^{\uparrow}$) of the spin-up bands are indicated next to each band ($C^{\downarrow}=-C^{\uparrow}$). The additional edge modes (indicated in purple) are not topologically protected. The highest energy edge-channel is localized to a single site along one direction, while the middle and lowest edge-channels are localized to two and three sites respectively. c. Experimentally observed coupling between two sites on the edge of the circuit, showing transmission through edge modes even within the bulk gaps. The structure within the gaps is to be expected, as the system is finite so the edge exhibits Fabry-P\'{e}rot resonances. d. Experimental site-resolved response to excitation at a central lattice site, within the band gap at 160kHz, showing bulk localization. e. Experimentally observed, site-resolved edge-mode structure at 160 kHz. The arrows in a. and c. reflect the frequency of excitation in b. and d.}
\end{figure}

The smoking gun of a topological band-structure is a gap in the bulk density of states within which spin-filtered edge-states reside~\cite{Hasan2010a,PhysRevLett.95.226801}. To probe this physics directly we excite the bulk of our meta-material (by driving an on-site inductor with a near-field coupled coil), and observe the response at other sites within the bulk (using a pickup coil). The resulting spectrum probes the RF density of states plus overlap factors reflecting the spatial mode profile at the drive- and measurement- locations. Figure \ref{fig2}a shows a typical measurement, with the predicted locations of the bands overlaid in orange; a gap is clearly apparent in the data. In Figure~\ref{fig2}d, we excite a central site at an energy between the bands and perform site-resolved microscopy. The insulating nature of the bulk is revealed by the exponential localization of the response. 

In Figure \ref{fig2}c, we excite and probe the system on its edge. We observe a response within the energetic band gap, experimentally confirming the existence of mid-gap edge modes, and strongly suggesting their topological nature, which we confirm below. It bears mentioning that in spite of the absence of a Fermi sea (our excitations are photons, which are bosons), we still observe an RF- insulating bulk and a conducting edge; this is because we are directly probing the density of states of the system at the RF-drive frequency. Furthermore, the appearance of clear Fabry-P\'{e}rot resonances on the edge, rather than a continuous density of states, is indicative of the periodic boundary conditions imposed by the closed edge. Figure \ref{fig2}e shows a site-resolved image of the edge mode at 160 kHz, demonstrating the persistence of the edge transport.

\section{Time-Resolved Dynamics}
\begin{figure}[h!]
\includegraphics[width=3.4in]{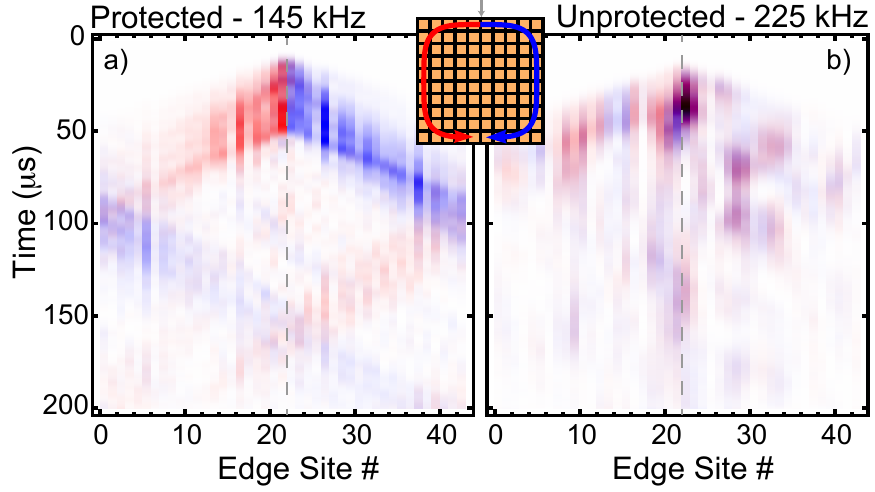}
\caption{\label{fig3}
Time-Resolved Transport Dynamics of the Edge Modes. a. Spin-resolved detection of the splitting of a localized $\textbf{A}$-site~$= (\uparrow + \downarrow)/\sqrt{2}$ excitation in the protected gap (at 145 kHz).  The $\uparrow$ (red) and $\downarrow$ (blue) signals are overlaid, demonstrating spin-momentum locking. Two round-trips are visible; sites 0 and 42 are equivalent. b. Spin-resolved excitation after exciting the same site in the unprotected upper edge mode at 225 kHz, where disorder immediately leads to backscattering for both spin-states. Inset: Edge lattice-site numbering convention.
}
\end{figure}

To demonstrate that the edge states are spin-orbit locked, we rapidly excite the $A$ inductor on a single edge site at an energy within the highest energy gap, and perform a spin-resolved time-domain measurement of the propagating excitation. Because $A=(\uparrow+\downarrow)/\sqrt{2}$, the excitation splits, with the $\uparrow$ ($\downarrow$) component propagating left (right).

Figure \ref{fig3}a shows the intensity at each lattice site around the system perimeter (yellow inset), as a function of time (increasing downwards), with red (blue) color channel indicating up (down) spin states. The splitting of the excitation, with red (blue) component moving left (right), demonstrates the spin-orbit locking of the edge states. The measured propagation velocity of 4.3(3) sites/10$\mu$s is consistent with the predicted group velocity of 4.2(2) sites/10$\mu$s arising from the numerical band-structure calculation (see SI). 

In contrast to the topologically protected edge modes between the third and fourth bands, the unprotected modes above the top band exhibit left- and right- propagating components for each of the top- and bottom- edges, for each spin state. Figure \ref{fig3}b shows the same dynamics experiment as in Figure \ref{fig3}a, but this time exciting the unprotected edge modes. Not only do both up- and down- excitations propagate both left- and right-wards, but both are rapidly backscattered and localized by disorder. These unprotected, termination-dependent edge channels are similar to zig-zag edges of a graphene nano-ribbon ~\cite{PhysRevB.54.17954}.

\section{Realizing a Spin-Orbit Coupled M\"{o}bius Strip}
 \begin{figure}[h!]
\includegraphics[width=3.4in]{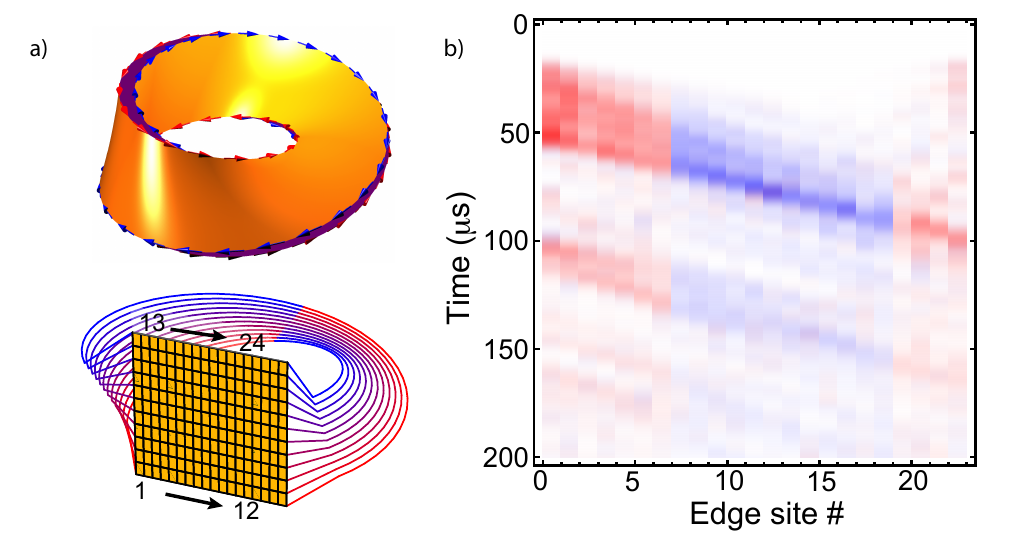}
\caption{\label{fig4}
Topological Insulator with M\"{o}bius Global Topology. a. Schematic M\"{o}bius topological insulator. Top: M\"{o}bius TI, with arrow indicating the edge propagation direction, and color the spin state. Bottom: Connectivity of the Mobius TI. PCB is shown in orange, with external connections generating the topology indicating spin on traversing the edge. This circuit is excited in the edge-channel at 145kHz, within the largest bulk gap, and the subsequent evolution of the wave-packet around the full perimeter of the system observed through time. b. Spin-resolved detection of edge-transport after excitation of $\uparrow$; $\uparrow$ and $\downarrow$ intensities are plotted in the red and blue channels, respectively, showing the conversion from $\uparrow$ to $\downarrow$ when the excitation moves from one edge to the other. Three round-trips are visible. 
}
\end{figure}
Because our topological meta-material is composed of lumped elements much smaller than a wavelength, it is possible to achieve exotic global topologies via zero-phase edge-to-edge connections that are difficult or impossible to realize with conventional materials. These new topologies have edges whose properties may be probed through time-resolved dynamics in our circuit. To achieve these new global topologies we add carefully designed capacitive couplings between the edges of the meta-material (see methods).

The M\"{o}bius topology is realized by connecting the left and right edges of the system with a half-twist and a spin-flip (Figure \ref{fig4}a). More precisely, the $A$ inductor $j$ sites from the top of the left edge is connected to the $B$ inductor at the $j$ sites from the bottom of the right edge. The spin flip is necessary, as ``up'' spins propagate rightwards on the top edge, but leftwards on the bottom edge~\cite{PhysRevB.84.193106}. Figure \ref{fig4}b shows the temporal dynamics of a spin up excitation in the system; it propagates rightward along the system edge, until it reaches the other edge, is converted into an down-spin, and continues its progression around the system perimeter; this repeats until the excitation is damped out by the finite system $Q$.

\section{Conclusion}
In this work, we have realized a topologically non-trivial band structure in a radio-frequency  circuit. Using site-resolved spectroscopy, we demonstrated the presence of bulk band gaps containing localized edge modes. Spin-resolved measurements in the time domain confirmed the spin-orbit locked character of the topologically protected edge modes, and enabled us to distinguish between protected edge modes within the gap and unprotected edge modes above the top bulk band~\ref{fig2}b. This approach opens a broad array of directions in the design and benchmarking of novel material structures, including long-range hopping to engineer truly flat bands~\cite{EliotKapit2010, PhysRevLett.106.236803}, and multi-orbital Chern insulators~\cite{PhysRevB.86.241112}. Extensions of the M\"{o}bius topology will enable studies of topological defects in layered spin-Hall systems~\cite{Ran2009}. In conjunction with superconducting quantum circuitry techniques~\cite{Wallraff2004}, the approach demonstrated in this work enables access to topological many-body phenomena~\cite{Hafezi2013} in the nearly-flat bands of the circuit, from anyons to fractional Chern insulators and beyond. 

\begin{acknowledgments}
We would like to thank C. Chin, M. Levin, J. Moore, and T. Witten for discussions. This work was supported by grants from the Air Force Office of Scientific Research, DARPA, and the University of Chicago MRSEC. \end{acknowledgments}

\section{Methods}
The topological meta-materials are composed of a room-temperature FR4 PCB, populated with 330pF capacitors with component-to-component scatter of $\pm1\%$, and 3.3mH wire-wound inductors with scatter of $\pm10\%$; this amounts to both spin-dependent and spin-independent disorder, with the former arising primarily from differential variation between the two inductors at a lattice site, and the spin-independent disorder from common-mode variation between lattice-sites. All disorder may be decreased by employing stripline inductors and capacitor, and micro-fabrication technology. In this case inductors and capacitors may be swapped, to move from left- to right- handed transmission lines. Because of the reduced bandwidth in the massive model (see SI), the sensitivity of the meta-material to component-scatter would be worse for a model with resonators on each site rather than just inductors- this corresponds to the difference between massive- and light-like lattice photons.

To inject photons into a site, we inductively couple to a single inductor by placing a 5 turn, 8mm diameter drive coil over the onsite inductor that we wish to excite. This inserts photons in the $A$ or $B$ sites; to inject spin-polarized $\uparrow$ or $\downarrow$ photons for Figure \ref{fig4}, the two inductors on a site are simultaneously excited $90^\circ$ out of phase by driving them with phase-locked synthesizers. Differential inductive in-coupling is compensated for by adjusting drive voltages on the two inductors. Spin-resolved detection is achieved by separate time-resolved detection of $A$ and $B$ sites, and transforming to the $\uparrow$ and $\downarrow$ states.

The M\"{o}bius topology is generated by employing a second PCB that provides the appropriate connections between the edges of the primary circuit TI PCB.

Numerical prediction of the steady-state frequency response is achieved by generating the admittance matrix of the circuit network and inverting it to compute the response to a drive. To calculate the band-structure, a coupled first-order system of differential equations representing the inductor-capacitor network was generated for a $1D$ strip 23 sites wide with fixed boundaries in the transverse direction, and periodic boundary conditions in the longitudinal direction, whose phase is set by the quasi-momentum $q$ under consideration. The system of equations was numerically diagonalized with eigenvalues correspond to the system energies at the given quasi-momentum (see supplement for details).

\newpage\newpage\newpage\newpage

\onecolumngrid
\appendix{}
\section{\textbf{Supplementary Materials}}
This supplement details the basic ideas behind creating an RF analog of a topological insulator.  
Because the RF lattice studied in this problem does not correspond to a tight-binding model, generating the band-structure and response functions is a bit more involved. What we have discovered in this work is that while our final circuit does not have a direct analog in other material systems, the essential feature, the local permutation of connections, results in a band-structure and dynamics that exhibit all of the signatures (both formally and experimentally) of a topological lattice model.

In section 1 we will describe, in a step by step manner, the transition from the theoretical, spinless, tight-binding  Hofstadter model to the experimentally realized, spinful, network model. The changes that will have to be understood are: (1) Controlling the sign of the coupling by employing an un-grounded network topology, (2) Leaving the limit of weak inter-site coupling thereby coupling positive- and negative- energy excitations, (3)Employing strong inductive couplings vs. strong capacitive couplings. In section 2 we will introduce the computational tools that permit calculation of band-structures, edge modes, and Chern numbers. These tools employ network admittance tensors, and transcend the limit of a tight-binding Hamiltonian. 

\section{From Harper-Hofstadter to Lattice Spin-Hall}

\begin{figure}[h!]
\includegraphics[width=4.4in]{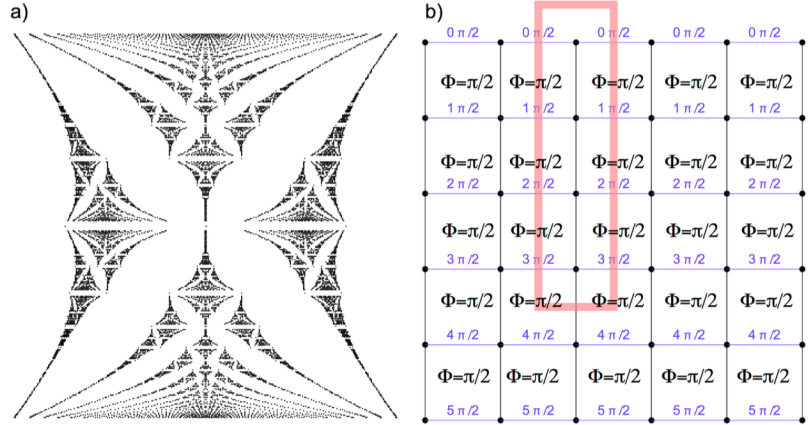}
\caption{\label{fig:Hofstadter}
(a) Hofstadter's butterfly: This figure shows the spectrum of allowed energies (vertical axis) versus magnetic flux per plaquette (horizontal axis), for a square lattice with uniform flux. (b) A particular gauge-choice for the Harper-Hofstadter model at a flux per plaquette of $1/4$, corresponding to a geometric phase of $\pi/2$ acquired for tunneling counter-clockwise around a single plaquette. In the chosen gauge, all vertical tunneling has phase zero, and horizontal tunneling accrues of phase of $j \pi/2$, where $j$ is the row. Because this phase is periodic every four sites, the new expanded unit cell, shown in red, is four sites tall.
}
\end{figure}

\noindent\emph{The Harper-Hofstadter Model, and Breaking Time-Reversal Symmetry}\\
When a massive particle is placed in a periodic 2D potential (lattice), its energy-momentum dispersion is modified by the lattice, forming bands of allowed and forbidden energies. Similarly, when a massive charged particle is placed in a uniform magnetic field, flat bands of allowed and disallowed energies, called Landau-levels, are formed. When both a lattice and a magnetic field are applied to a massive charged particle, the bands induced by the magnetic field and lattice compete- the resulting fractal band structure, as a function of magnetic field strength and energy, is known as the Hofstadter butterfly, and is shown in (SI Figure~\ref{fig:Hofstadter}). The basic story is that if the magnetic field induces free-space cyclotron orbits whose area is a fully reduced rational P/Q lattice cells, then the band structure exhibits Q distinct bands (SI Figure~\ref{fig:Hofstadter}). Equivalently, in the presence of a magnetic field, the system becomes periodic every Q unit cells~\cite{Hofstadter1976}. This may be seen by writing out the tight-binding Hamiltonian, using the Peierl's substitution~\cite{Hofstadter1976}:

\begin{dmath}
H_{\rm Hofstadter} = -t \sum_{m,n} a_{n,m}  \left( a_{n,m+1}^\dag + a_{n,m-1}^\dag  + a_{n-1,m}^\dag e^{i m \phi} + a_{n+1,m} ^\dag e^{-i m \phi} \right)
\end{dmath}

where $t$ is the nearest neighbor tunneling rate, and $\phi = 2\pi P/Q$ is the magnetic flux per plaquette. It is clear that the new magnetic unit cell is $Q$ lattice sites tall, as the exponential factor repeats ever Q lattice sites.
The difficulty in implementing this Hamiltonian for uncharged particles, which thus do not respond directly to magnetic fields via a Lorentz force, is that $H_{\rm Hofstadter} $ is not invariant under time reversal. This can be seen explicitly by verifying that the complex conjugate of $H_{\rm Hofstadter} $ is not the same as $H_{\rm Hofstadter} $ due to the complex exponential terms. Such a time-reversal asymmetric terms require something beyond inductors and capacitors (or equivalently, waveguides) to implement; possible approaches include periodic modulation~\cite{Jotzu2014,Lindner2011,Wang2013,Rechtsman2013a,Kitagawa2010} and coupling to materials with handed susceptibility induced by real magnetic fields~\cite{Wang2008}.

\noindent\emph{Using Spin-Orbit coupling to recover Time-Reversal Symmetry}\\
The simplest way to recover time-reversal symmetry is to add a second spin state that experiences the opposite magnetic field. Such physics typically arises due to Rashba or Dresselhaus spin-orbit coupling, after an appropriate gauge transformation and Peierls substitution~\cite{Hofstadter1976}:

\begin{dmath}
H_{\rm SOC} = -t \sum_{m,n} a_{n,m} (a_{n,m+1}^\dag + a_{n,m-1}^\dag +a_{n-1,m}^\dag e^{i m \phi S_z} + a_{n+1,m}^\dag e^{-i m \phi S_z}) $$
\end{dmath}

The key point is that in the spinor model, the anti-unitary time-reversal operator not only takes a complex conjugate of the Hamiltonian, but also flips spins. Under such an operation it is clear that $H_{\rm SOC}$ is time-reversal invariant.

What this does not yet tell us is how to engineer a spin that flips under time-reversal. It turns out that for any system of two degenerate energy levels, there is a basis $(0,1)$ for which the basis-states are invariant under time-reversal, and a basis $(\uparrow,\downarrow)$ for which they swap.

The above can easily be seen using our experimentally realized system as an example: consider an excitation that may live in either of two resonant circuits A or B, as in the main text. In this case, reversing time certainly does not move the excitation from resonator A to resonator B. Thus the basis (0,1)=(A,B). By contrast, if we choose an excitation that is in a rotating superposition of the two resonators, $(A \pm iB)/\sqrt{2}$, reversing time corresponds to taking a complex conjugate or equivalently reversing the rotation direction, and thus swaps the excitation. Thus $(\uparrow,\downarrow)=(A \pm iB)/\sqrt{2}$. This may also be seen by noting that the time-reversal operator is a unitary matrix times a complex conjugation operator~\cite{Shankar2008} .

What we have shown is that a pseudo-spin degree of freedom may be engineered using simple, time-reversal-invariant linear components; the spin ``flips'' under time-reversal in an appropriately chosen basis. What remains, then, is to engineer an effective magnetic field that acts oppositely on the two spin states. All we have shown so far is that this should be possible without violating time-reversal symmetry, as the entire target system is time-reversal invariant.

Put another way- in the basis of the up- and down- pointing spins, the Hamiltonian will supply a time-reversal-symmetry breaking magnetic field to each spin state, corresponding to a Lorentz force. However, because the up- and down- pointing spin states themselves are time-reversal breaking quantities, the whole Hamiltonian does not violate time-reversal symmetry. This hints at the possibility perhaps such a Hamiltonian can be engineered from entirely linear, reciprocal components.

\section{Experimental Realization}
To gain a quantitative understanding of the behaviors of a photon in a topologically non-trivial microwave circuit, it is instructive to return to the simple case of a lumped-element 1D transmission line and its generalizations.\\

\noindent\emph{Engineering Photon Dispersion in 1D Microwave/RF Lattices}\\

\begin{figure}[h!]
\includegraphics[width=3.4in]{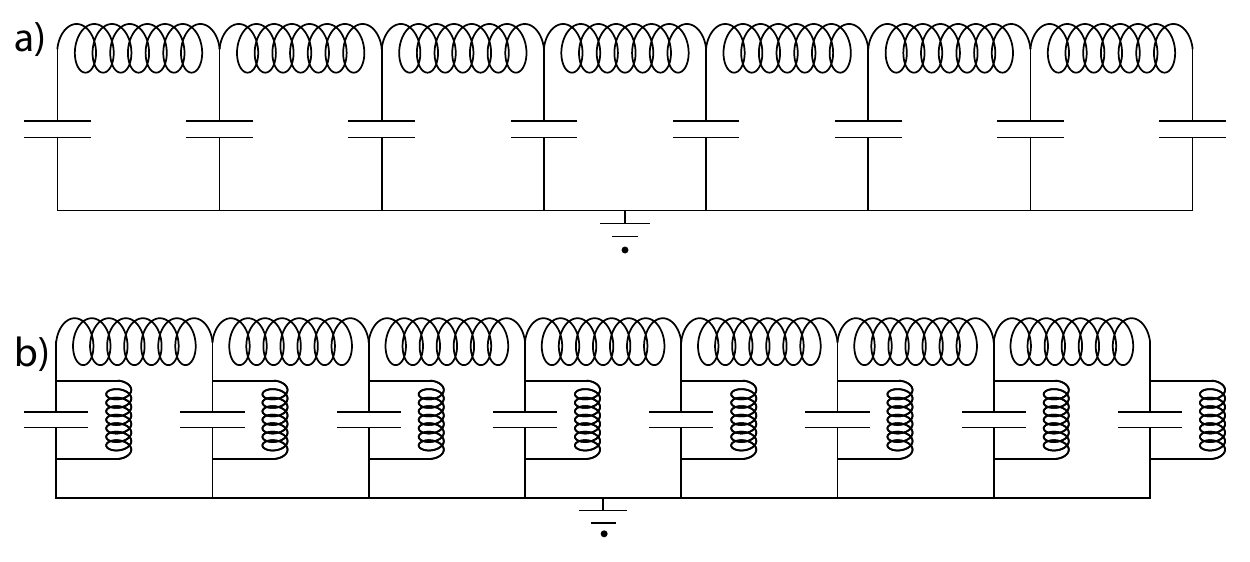}
\caption{\label{fig:lumped_txline}
(a) A grounded lumped-element model of a  transmission line, with light-like dispersion at low-momenta, and a cutoff at a quasi-momentum given by the inverse of the inter-site spacing. (b) A grounded lumped-element model of a tight-binding lattice.
}
\end{figure}

\noindent\emph{1D Transmission Line:} Light in a transmission line has a linear dispersion, reflecting the fact that information travels down the transmission line with a constant velocity $v=c/n$, where $c$ is the speed of light in vacuum and $n$ is the index of refraction of the transmission line. Such a system may be modeled by a capacitors which are coupled to one another by inductors, where the capacitor value is the given capacitance per unit length of the transmission line, and the inductor value is given by the inductance per unit length of the transmission line. We consider the more familiar case of sites that are grounded on one side (SI Figure~\ref{fig:lumped_txline}). This lumped-element model has a linear dispersion at low momentum, reflecting the behavior of its continuum counterpart. As the momentum approaches the wave-vector of the lumped elements, the dispersion levels off: $\omega =2 \omega_0 \sin(q/2)$, with $\omega_0^2=1 / L C$, with $L$ and $C$ the lumped inductance and capacitance respectively, and $q$ the quasi-momentum in units of inverse lumped element sites.

\noindent\emph{1D Tight-Binding Lattice for Photons:}
To make the photons appear ``massive,'' an inductor $L_{\rm site} \ll L$ may be added in parallel with the on-site capacitors. In this case the system behaves not as a transmission line, but as a string of coupled resonators. These resonators have a ``rest-mass'' of $\omega_{\rm res}^2=1/(L_{\rm site} C)$ that rounds out the dispersion at low quasi-momenta, imbuing the photons with a mass:

\begin{dmath}
\omega=2 \omega_0 \sqrt{\frac{L}{4L_{\rm site} } + \sin^2(q/2) } \approx ≈2\omega_0 \left(\frac{L}{4 L_{\rm site}}\right)^{1/2} (1+ \frac{L_{\rm site}}{L} \sin^2(q/2))
\end{dmath}
 in the limit $L_{\rm site} \ll L$.

The width of the band is then $4 \omega_0 (L_{\rm site}/4L)^{1/2}$. In a tight binding lattice the bandwidth is $4 t$, where $t$ is the nearest-neighbor tunneling rate. Thus we can now identify the tunneling rate as $t\approx \omega_0  (L_{\rm site} / 4 L)^{1/2}$.\\

\noindent\emph{Calculational Tools:}
The above expressions were computed from a more general frequency-domain approach, assuming a (frequency dependent) on-site impedance $Z_{\rm site}$ to ground, and a coupling impedance between sites $Z_{\rm couple}$. From here the Kirchhoff's laws may be applied to the on-site voltages and inter-site currents to produce the following expression implicitly relating the frequency to the quasi-momentum: $4 \sin^2(q/2)=-Z_{\rm couple}(\omega)/Z_{\rm site} (\omega)$.\\

\noindent\emph{Right- vs Left- Handed Circuits:}
With this more general frequency-dependent expression in hand, it is clear that swapping inductors and capacitors in these circuit networks is equivalent to sending $\omega \rightarrow \omega_0^2/\omega$ in the dispersion relation; such left handed transmission lines are well-studied~\cite{Caloz2004}, and their extension to two-dimension will be employed later in this work, in our discussion of the full topologically insulating circuit.\\

\begin{figure}[h!]
\includegraphics[width=3.4in]{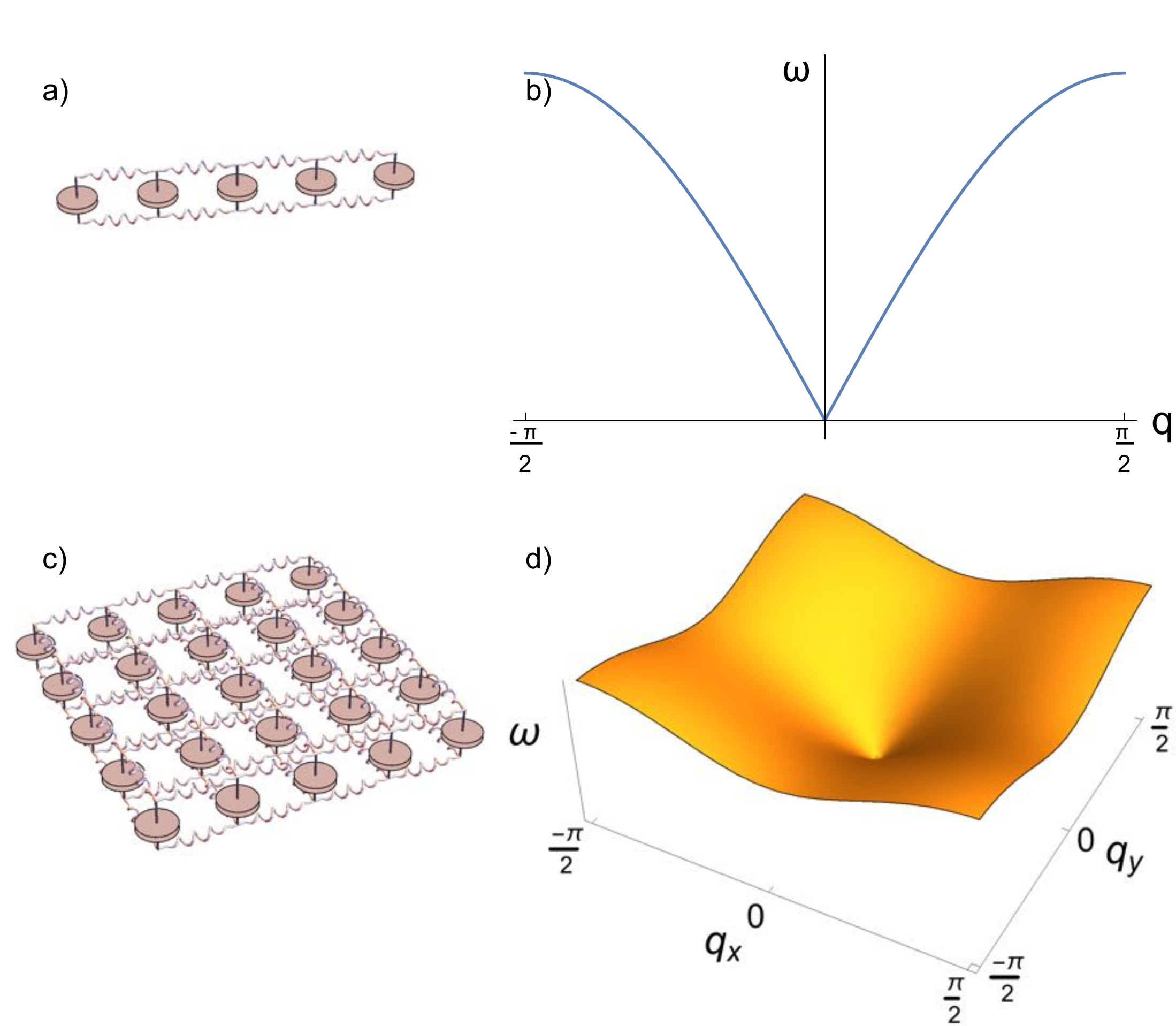}
\caption{\label{fig:1D2D}
(a) In the lumped-element model of a transmission line, the capacitance per unit length is modeled with an actual capacitor, and the inductance with an actual inductor. (b) This system has a linear (light-like) dispersion at low momenta, rolling off at a momentum given by the spacing of the lumped-elements. This roll-off reflects the fact that a the lumped-element system behaves differently the continuous transmission line once the dynamics probe the discreteness of the lumped-element system (absent in the continuous system). (c) The two-dimensional extension of this model exhibits (conical) light-like 2D dispersion (d). For symmetry with what follows, we draw double-ended connections between the capacitors; the calculations in this section assume that the lower-side of the capacitors are grounded.
}
\end{figure}

\noindent\emph{Grounded vs Symmetrical Networks: Zero Frequency Modes.}
It also bears mentioning that up to this point we have treated the on-site impedance network as grounded on one side; SI Figure~\ref{fig:1D2D} depicts the circuit symmetrically, without grounding, and indeed such a setup results in quantitatively identical behavior once $Z_{\rm site}$ is replaced by $Z_{\rm site}/2$: The symmetrical case admits both even and odd solutions: the even solutions are all at zero frequency, corresponding to a DC shift at various quasi-momenta, and hence may be ignored, while the odd solutions are the same as those reached in the grounded circuit network. The symmetrical generalization will become important in the 2D system, where the sign of a site-to-site coupling may be reversed (made negative) by connecting the positive end of one site to the negative of the other, and vice-versa; in 1D such a choice can always be gauged away; not so in 2D: this is the origin of the spin-hall effect in our lattices.\\

\noindent\emph{Engineering Photon Dispersion in 2D Microwave/RF Lattices}\\
In this section we will extend from 1D to 2D transmission lines, and add a spin degree of freedom, and finally a synthetic spin-orbit coupling. We will describe, along the way, the calculational tools necessary to compute 1)the modes of the infinite system, 2)the edge states in a strip with periodic boundary conditions at finite quasi-momentum and finally 3)the two-point couplings in a finite lattice.

A 2D microwave circuit:
The direct extension of the transmission line to two dimensions is a square lattice of grounded capacitors connected to their nearest neighbors with inductors, as shown in the lower left panel of SI figure~\ref{fig:1D2D}. The dispersion for such a system, shown in SI figure~\ref{fig:1D2D}d, is given in general by:
$4 (\sin^2 (q_x / 2)+\sin^2 (q_y / 2)) =- Z_{\rm couple}(\omega)/Z_{\rm site} (\omega)$, or $\omega \cong \omega_0 |q|$, for low momenta $q \ll \pi$ in the special case of on-site capacitors and coupling inductors. This dispersion corresponds to the dynamics of waves propagating in the plane transverse to the two plates of a parallel plate capacitor—in short, the 2D analog of a waveguide or transmission line. Note that it is only isotropic at low momenta, where the sines can be expanded to linear order in their argument. 

Again we can make the photons appear massive (both in the lowest allowed energy, and in their dispersion/dynamics) by adding an on-site inductor, and again we can compute the dispersion of the system with inductors and capacitors swapped by sending $\omega \rightarrow \omega_0^2/ \omega$.\\

\noindent\emph{Adding a spin degree of freedom:}
To add a spin-degree of freedom to the system, we simply make a second copy of the 2D lattice structure: now there are two capacitors on each site, which we will label A and B or red and blue, and initially A capacitors are coupled to A capacitors by inductors, and B capacitors to B capacitors by another set of inductors and no A-B couplings, as in SI figure~\ref{fig:1D2D}. Because the A and B sublattices are uncoupled, the dispersion relation is the same as a single-spin-state 2D lattice, with a two-fold degeneracy.
 
 \begin{figure}[h!]
\includegraphics[width=3.4in]{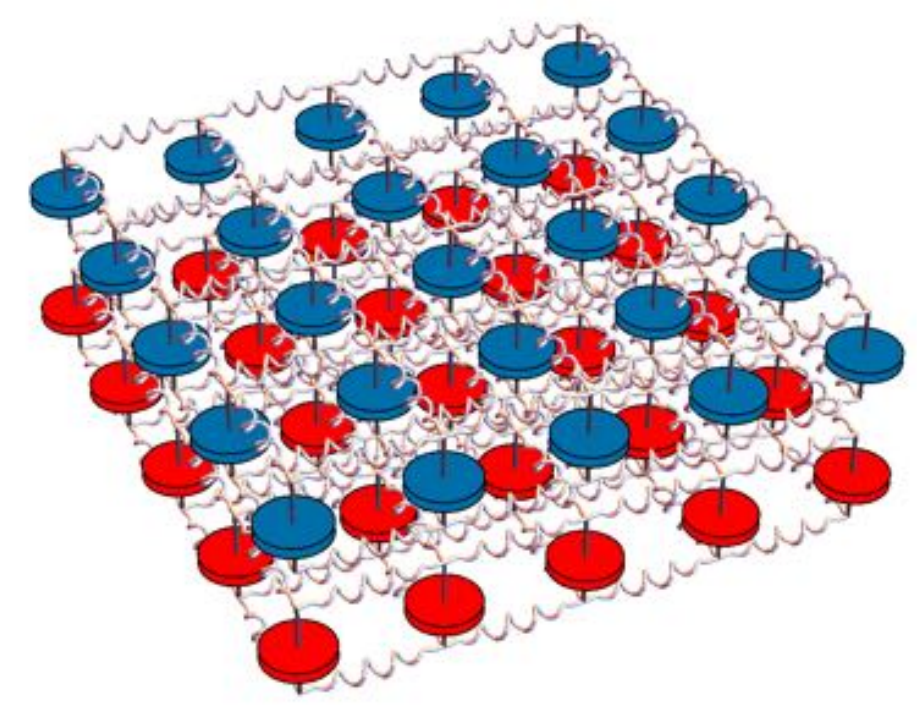}
\caption{\label{fig:spinlattice}
Two-spin 2D photonic waveguide realized with discrete components. A photon in state A lives in the red lattice of capacitors, and a photon in state B lives in the blue lattice. The A- and B- lattices are uncoupled from one another.
}
\end{figure}

\noindent\emph{Adding a synthetic spin-orbit coupling:}
To add the spin-orbit coupling, we must couple the A and B sublattices. As discussed in section 1 of this supplement, the states which map onto one-another under time reversal are $(\uparrow,\downarrow) = (A \pm iB)/\sqrt{2}$, and so it is these states which must experience synthetic gauge fields of opposite sign.

For simplicity, we will consider first the $\uparrow$ state, with a flux per unit cell of $\alpha=1/4$: The Peierl's substitution dictates that we need to generate a tight-binding horizontal tunneling with a phase of $j × \pi/2$, where $j$ is the current row of the system. In the case of a ``massless'' photonic lattice, ``tunneling'' is not well-defined (the dispersion being linear at low momenta, rather than sinusoidal), so we will, for now, consider the massive case with the onsite inductor on each lattice site, and later send this onsite inductance to infinity.

A tunneling phase of $\pi/2$ sends the state $\uparrow = (A+iB)/\sqrt{2}$ on site 1 to the state: $e^{i \pi /2} (A+iB)/\sqrt{2} =(iA-B)/\sqrt{2}$ on site 2: in other words, the A resonator on site 1 needs to be inductively coupled to the B resonator on site 2 with a minus sign, while the B resonator on site 1 needs to be inductively coupled to the A resonator on site 2 without a minus sign. The beauty here is that we never need to explicitly generate a complex tunneling matrix element (which we cannot do, as we cannot break time reversal with only lossless inductors and capacitors) as we are only transferring existing $i$'s between A and B resonators on adjacent sites: this is the essence of time-reversal symmetric dynamics.

 \begin{figure}[h!]
\includegraphics[width=3.4in]{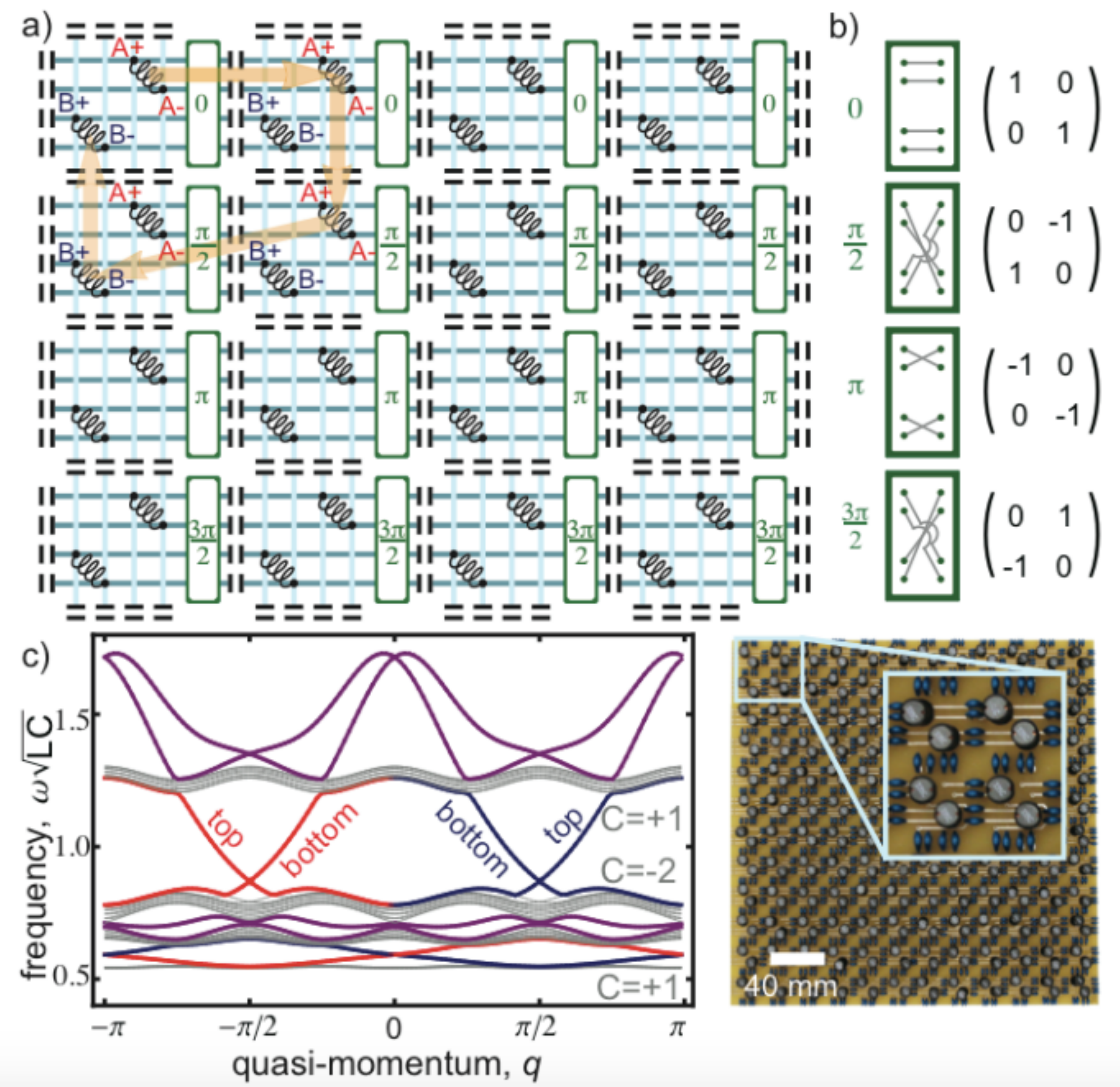}
\caption{\label{fig:circuitmodel}
(a) Circuit topological insulator schematic. The periodic structure is formed by onsite inductors and coupling capacitors, (black) that are connected via a latticework of wires (light and dark blue lines). At each lattice site, the two inductors ``A'' and ``B'' correspond to right and left circularly polarized spins. When a photon traverses a single plaquette (indicated in orange) it accumulates a Berry phase of $\pi / 2$. The phase is induced by braiding (indicated by the green boxes and specified in b) of the capacitive couplings. (b) Structure of the coupling elements between lattice sites. Each row shows one of the four rotation angles implemented by the capacitive coupling in the circuit. The rotation angle (left column) is induced by connecting inductors as shown (middle column). The corresponding rotation matrices (right column) indicate the inductors being coupled, as well as the signs of the couplings. (c) Band Structure of a Circuit TI. A strip of circuit TI with fixed boundary conditions in the transverse direction is numerically diagonalized at finite longitudinal quasi-momentum q, yielding massive bulk bands (gray), and spin-orbit-locked edge states (red=$\uparrow$, blue=$\downarrow$) that reside in the bulk gap. Labels of ``top'' and ``bottom'' denote the boundary that each edge mode propagates along (with opposing group velocities d$\omega$/dk). The Chern numbers of the spin-up bands are indicated next to each band. The additional edge modes (indicated in purple) are not topologically protected. The highest energy edge-channel is localized to a single site along one direction, while the middle and lowest edge-channels are localized to two and three sites respectively. (d) Photograph of Circuit Topological Insulator. The inductors (black cylinders) are coupled via the capacitors (blue); circuit topology is determined by the trace layout on the PCB (yellow). Inset: Zoom-in on a single plaquette consisting of four adjacent lattice sites.
}
\end{figure}

To realize the necessary negative tunneling matrix elements, it is be possible over a narrow bandwidth to couple with a capacitor instead of an inductor, thereby inverting the sign of the impedance. A more robust, broadband solution is to employ a double-ended, symmetric circuit topology, and inductively couple the top side of one resonator to the bottom side of the adjacent resonator. This approach is essential as we work in a maximally broadband configuration, by removing on-site inductors. As we will shortly show, this does not impact band topology. Finally we invert the circuit topology, swapping inductors and capacitors and sending $\omega \rightarrow \omega_0^2/\omega$. This is important, practically, to minimize the number of bulky inductors compared with the much smaller capacitors, and also does not impact the band topology. The circuit that was ultimately implemented is shown in SI figure~\ref{fig:circuitmodel}.\\

\noindent\emph{Calculating Two-Point Transport Functions for a Finite System}
All of our numerics are based upon a frequency domain approach. We compute an admittance matrix $Y_{i j}=1/Z_{i j}$, where $Z_{i j}$ for $i \neq j$ is the direct impedance between nodes $i$ and $j$. and $Z_{ii}$ is the impedance between node $i$ and gnd. It is very important to note that $i$ and $j$ are not spatial indices for an $N$ by $M$ lattice, $i$ and $j$ must each must take on $4NM$ unique values, corresponding to the $N×M$ lattice sites, with the two ends of each of two (A and B) inductors on each lattice site. The admittance matrix is thus $4NM$ by $4NM$.

Subject to a source current distribution $I_i^{\rm drive} (\omega)$, Kirchhoff's laws for the voltage at node $j$, $V_j (\omega)$, provide $\sum_j Y_{ij} (\omega) V_j (\omega)=I_i^{\rm drive} (\omega)$. We can thus solve for the voltage distribution according to: $V_i (\omega)=\sum_j (Y(\omega) ^{-1})_{ij} I_j^{\rm drive} (\omega)$.

In practice, we (inductively) inject a current at a single site $\mu$, $I_j^{\rm drive}=I_0 \delta_{j,\mu}$, and (inductively) measure the voltage at site $\nu$, $V_{\nu} (\omega)=I_0 (Y(\omega)^{-1} )_{\mu \nu}$. Thus the two-point transport information is entirely contained in the matrix elements of admittance tensor.

The ungrounded, symmetrical nature of the circuit results in infinite-frequency resonances of the admittance tensor preventing it from being inverted. While we could simply invert in the orthogonal subspace, in practice it is simple to include a vanishingly small capacitance from every node of the circuit to ground (which realistically arises from stray capacitances)- this moves the poles away from infinite-frequency, allowing inversion of the matrix; these high-frequency modes do not quantitatively impact the behavior of the circuit.\\

\noindent\emph{Calculating Band-Structure in the Spin-Orbit Coupled Lattice}\\

 \begin{figure}[h!]
\includegraphics[width=3.4in]{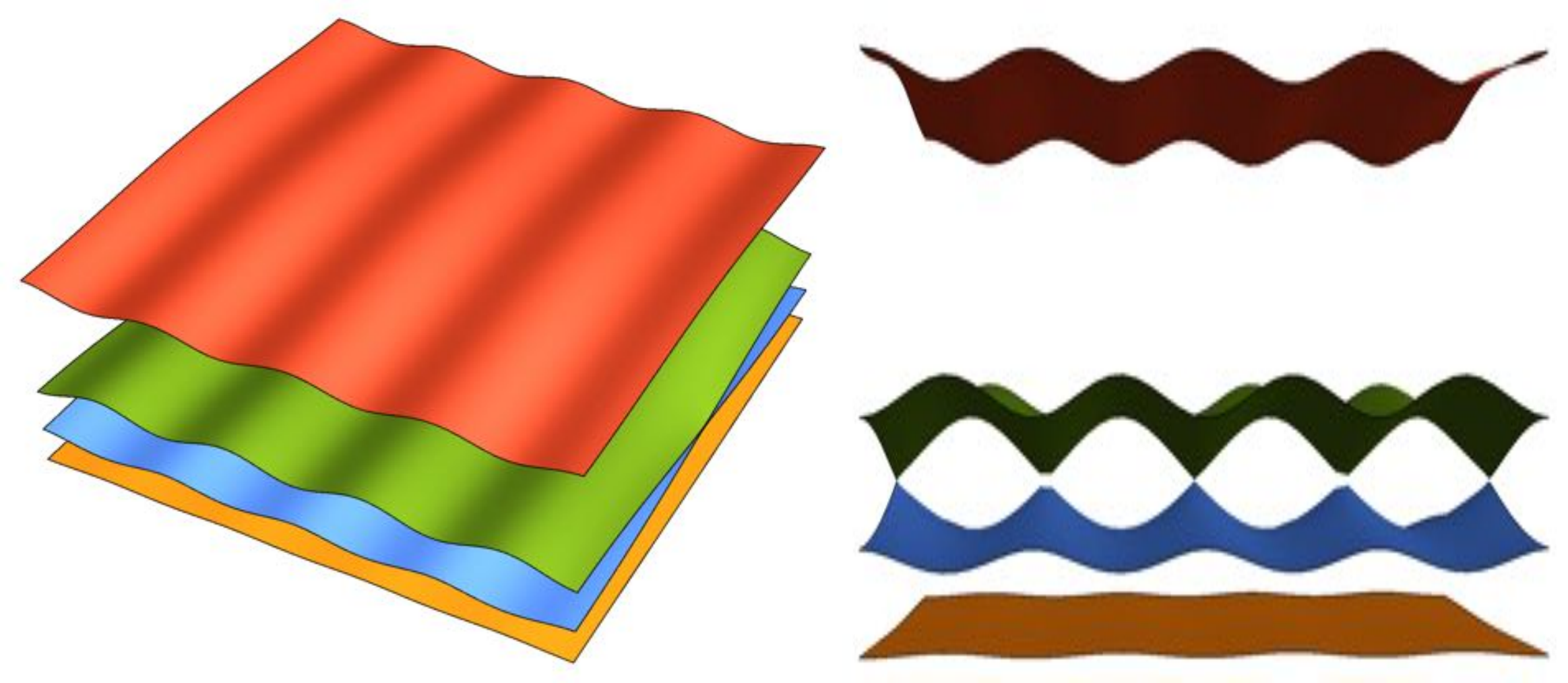}
\caption{\label{fig:Bandstructure}
Band structure for the topologically insulating circuits from two different perspectives (left/right), showing four distinct bands, with the middle two touching at four Dirac points. Note that each band is two-fold degenerate, for up- and down- spins.
}
\end{figure}

To compute the band-structure of the lattice, we consider a single ``extended'' (magnetic) unit cell containing a full flux quantum, which at a flux per plaquette of $1/4$, is a 1x4 strip of sites. We employ  a circuit analog of Bloch's theorem, which amounts to a wave-function ansatz which is identical from one extended unit cell to the next up to a factor of $e^{\pm i q_x }$ in the $\pm x$ direction, and $e^{\pm i q_y}$ in the $\pm y$ direction. We can then apply Kirchhoff's laws to generate an admittance matrix $Y_{ij} (\omega,q_x,q_y)$ as a function of frequency and quasi-momentum. The mode frequencies are the roots of the determinant of $Y$, but computing these is computationally taxing, and will not provide the mode functions, which we will require for computation of the Chern numbers of the bands. A more general approach is to generate the equations of motion from Y. To this end we write $i \omega Y$ as a Laurent series in $\omega$: 
$Y_{ij} = (1 / i \omega L)_{ij} + i \omega C_{ij}$
It should be apparent (from the fact that $Y$ gives the admittance between sites, which for a capacitor is $i \omega C$, and for an inductor is $1/i \omega L$) that C parameterizes the capacitive couplings, and L the inductive couplings.  We can thus write the equations of motion in the time domain as: $\sum_{j,k} L_{i j} C_{j k} \ddot{V}_k +V_j=0$

The mode frequencies are thus the square roots of the eigenvalues of the matrix $(LC)^{-1}$, and the bloch wavefunctions are the corresponding eigenvectors. The numerically calculated band energies are plotted vs quasi-momenta $q_x$ and $q_y$ in SI figure~\ref{fig:Bandstructure}.\\

\noindent\emph{Computing Band Chern Numbers:} To numerically compute the Chern numbers of each of the up- and down- spin bands, we employ a gauge-independent definition of the Chern number:
$$C^{\uparrow,\downarrow}_b = \frac{1}{2 \pi i} \int \int_{BZ} Tr\left(\rho^{\uparrow,\downarrow} \left[\frac{d \rho^{\uparrow,\downarrow}}{dq_x },\frac{d \rho^{\uparrow,\downarrow}}{dq_y } \right] \right) d^2 q$$
with $\rho^{\uparrow,\downarrow} (q)\equiv \sum_b \left(\psi_b^{\uparrow,\downarrow} (q)\right)^\dag \left(\psi_b^{\uparrow,\downarrow} (q)\right) $.

Here the sum over $b$ runs through the bands under consideration: For the top and bottom bands, we compute their Chern numbers $C_1$ and $C_4$ independently. Because the middle two bands (2 and 3) touch (at Dirac points), their Chern numbers cannot be computed independently, so we must compute $C_{2,3}$ as a single entity. The above definition of the Chern number follows from the standard definition in~\cite{Hasan2010a} after a bit of algebra. 

The $\psi_b^{\uparrow,\downarrow} (q)$ are the eigenvectors extracted from the band structure calculation in the previous section, with the added complication that we must project onto a single-spin subspace. To this end we compute eigenvectors and eigenvalues of $P^{\uparrow,\downarrow} (LC)^{-1} P^{\uparrow,\downarrow}$, with $P^{\uparrow,\downarrow} \equiv \frac{1}{2} \sum_l (A_l \pm i B_l )( A_l^\dag∓\mp i B_l^\dag )$   , the projector onto the up- (down-) spin subspace. Here the $A_l$ ($B_l$) are the state vectors corresponding to a voltage exclusively at the A (B) sub-lattice of site $l$. 

It is worth noting that we have intentionally avoided bra-ket notation, typically employed in defining and computing Chern numbers; this was done to emphasize that all of the topological properties of the bands studied here are classical- the physics applies equally well to individual particle quantum wave-functions as to classical fields.\\

\noindent\emph{Calculating the Mode Structure of an Infinite Strip:}
The full 2D band-structure of a 2D system cannot exhibit edge-states, as the system has no edges! To study the properties of the edge states, we investigate the mode-structure of a strip of finite width, but infinite length; the top and bottom of the strip will now support edge modes, at the cost of projecting the resulting band-structure into one dimension.

To compute the spectrum of modes of our infinite strip, we employ the tools of the preceding section, but instead of diagonalizing a single magnetic unit cell with Bloch's theorem applied along both $x$ and $y$, we only employ Bloch's theorem along $x$ and consider a single strip of $L$ unit cells along $y$ with fixed boundary conditions at the top and bottom, defining the location of the edge. In practice this means that adjacent rows are assumed to be identical to the row under consideration, up to a factor of $e^{i q_x}$.

The result of such a diagonalization, versus quasi-momentum $q_x$, is shown in SI Figure~\ref{fig:circuitmodel}c, with the bulk bands designated in gray and the edge states connecting them designated in red, blue, and purple. While edge states may exist in many systems, their existence is required if the sum of the Chern numbers of the bands below is non-zero (mod 2), and they are further guaranteed to be topologically protected, a hallmark of a topological insulator.

\bibliographystyle{naturemag}
\bibliography{TopologicalInsulator.bib}
\end{document}